\documentclass[a4paper,10pt]{article}

\usepackage[ansinew]{inputenc} 
\begin{document}



{\large {\bf  Hadron Spectroscopy (theory): \\
Diquarks, Tetraquarks, Pentaquarks and no quarks}}

 F.E. CLOSE\footnote{\tt{e-mail: f.close@physics.ox.ac.uk}} 

{\it Rudolf Peierls Centre for Theoretical Physics,
University of Oxford, Oxford OX1 3NP, United Kingdom}\\

\begin{abstract}
States beyond those expected in the simple constituent quark model are now emerging.
I focus on the scalar glueball and its mixing with states in the $q\bar{q}$ nonet, and also
on correlations in Strong QCD that may form diquarks and seed $qq\bar{q}\bar{q}$ states.
Some models of the pentaquark candidate $\Theta(1540)$ are critically discussed. 
\end{abstract}

{\bf The meson landscape}

This year we have seen several hadrons announced that do not fit easily with the simple
valence picture of $q\bar{q}$ or $qqq$ mesons and baryons. With hindsight one might wonder why
it took so long. This simple picture exploits degrees of freedom that transform like the
fields of $L_{QCD}$ but are not identical to them. Two quarks attract one another in $\bf{\bar{3}_c}$ 
with about the strength of $q\bar{q}$ coupled to colour singlet and so should play a significant role in
in generating the colour degrees of freedom in Strong QCD. For light flavours there is even an old
calculation\cite{voglweise} suggesting that
the effective mass of the antisymmetric $[ud]$ pair, ``scalar diquark",
 is comparable to that of a single $q \equiv u,d$. There is even some phenomenological support for 
 this\cite{sussp,wilc}.
 If so it is energetically as easy to make colour singlets
 from $[qq][\bar{q}\bar{q}]$ as from $q\bar{q}$.
 
 The low lying scalar mesons fit well with this idea\cite{maiani,Jaf77,CT02}. This strong attraction of flavour
 antisymmetric scalar diquarks should even imply exotic
 combinations made from $[cs][\bar{u}\bar{d}]$, $[cd][\bar{u}\bar{s}]$ and $[cu][\bar{u}\bar{s}]$\cite{Lip77}.
 The idea that baryons may emerge
 naturally as excitations of a quasi-two centred system has been resurrected\cite{wilc}. In turn this raises
 questions about other energetically favoured examples of such correlations. Two $[ud][ud]$ couple attractively to
 $\bf{3_c}$ (probably forced into $L=1$ by Bose symmetry\cite{JW})
 and so need a further $\bf{\bar{3}_c}$ to saturate. One way would be to add a third $[ud]$ (and 
 another L=1 but even so the diquark mass 
 cannot be too low if we are not to end up with a state more stable than the deuteron!) or a $\bar{q}$. If the
 latter is $\bar{s}$ we have a manifestly exotic strange baryon with the quantum numbers of the
 $\Theta$, evidence for and against which has been extensively reviewed here\cite{jinshan}.
 
 But why stop at the diquark? A $[ud\bar{s}]$ combination also is strongly attractive and with different flavours
 does not suffer annihilation via gluons. This enables one to construct the $\Theta$ quantum numbers with a
 quasi-two centred system, $[ud\bar{s}][ud]$ with $L=1$ needed to keep would be repulsive correlations 
 apart\cite{KL}. Completing the simple quasi-two centred states are attractive combinations such 
 as $[ud\bar{s}](\bar{s})$. The phenomenology of these includes flavour {\bf 10} and $\bf{\bar{10}}$ mesons,
 which might relate to exotic mesons with $J^{PC}=1^{-+}$ at 1.4-1.6GeV\cite{fcb}, but the detailed
 similarities and differences with the $[qq][\bar{q}\bar{q}]$ remains to be investigated.
 
 Most attention has focussed on the $\Theta$ and its implications for correlations as above. I shall
 not review this literature due to limitations of space and as it is well known, but I shall raise
 some questions that remain to be answered. While the above remarks may turn out to be critical in
 understanding the degrees of freedom for light flavours, the heavy flavours are better understood. Their
 phenomenology gives hints as to how to begin unravelling the code of the light flavoured sector.

So I shall begin with
the $b\bar{b}$ and $c\bar{c}$ traditional realm where the non relativistic model works well, 
at least as far as the S and P wave combinations are concerned. In particular note the scalar mesons
are canonical: they are in the right place, the E1 radiative transitions from $2^3S_1$ and their decays
to $1^3S_1$ appear to be in accord with theory\cite{pdg}. The $^1P_1$ charmonium state has been reported\cite{rosen}.

A novel entree to light hadrons is emerging with the decays of these $\chi$ states\cite{harris}. 
 $\chi_0 \to \pi^+\pi^-\pi^+\pi^-$ shows $f_0(980)f_0(980)$
pair production at a strength similar to that of $K\bar{K}$; this is a surprise if the
$f_0(980)$ is purely a $qq\bar{q}\bar{q}$ state and suggests that in {\bf production} by hard gluons
the ``simplest" $q\bar{q}$ or $gg$ configuration dominates the dynamics (I shall contrast this with its appearance
in $\phi \to \gamma f_0(980)$ later). In the present data $\chi_0 \to \pi +(3\pi)$ should be studied to see if the
$\pi(1800)$ is prominent: this is a potential candidate for a gluonic excitation of the $\pi$ and is degenerate
with the $D$. This is an example of how light hadron dynamics needs to be understood as it can affect $D$ decays:
the Cabibbo suppressed decays of the $D$ can be affected by mixing with this $\pi(1800)$\cite{fclipkin}.
We now await data on
$\chi_1 \to \pi + X$; this is intriguing because
 in S-wave $X \equiv 1^{-+}$, predicted to be the lightest
 exotic gluonic hybrid channel. Production by
the short distance gluons in $\chi$ decays is thus eagerly awaited.

What do we expect to find in the spectroscopy of light flavours? The $Q\bar{Q}$ pattern
of S P D states that was apparent for heavy quarkonium seems to
survive for light flavours, though there is no fundamental a priori 
reason why we should have expected this. There is however a clear indication of where it does not work.
In the P-wave states the $2^+,1^+$ nonets, each containing two isoscalars representing the
$s\bar{s}$ and $n\bar{n}$ flavour combinations, are clearly seen 
even though these states are now above threshold for decays into mesons. 
The rule seems to be that the spectroscopy seeded by a short range $q\bar{q}$ 
remains visible so long as 
there are no open {\bf S-wave} hadron channels, which obscure the underlying short range
structure. This is particularly obvious in the $0^+$ sector. 
 Above 1GeV we find three I=0 (1370;1500;1710) - or even a fourth
$f_0(1790)$\cite{jinshan} - in place of two, and
below 1 GeV there are certainly two further states $f_0(980)$ and $a_0(980)$ and attractive channels
hinting at a full nonet including a further I=0 $\sigma(600)$. Intriguingly, this proliferation is in
accord with simple ideas from QCD.

Above 1 GeV Lattice QCD predicts a scalar glueball, mass $\sim 1.6$ GeV\cite{glueball,CT96},
which mixes\cite{AC95} with the isoscalar $q\bar{q}$ in its vicinity. In the limit of large mixing, 
 the flavour eigenstates tend towards {\bf 1+G}, {\bf 8}, {\bf 1-G}\cite{CT96}. Fits to the
pseudoscalar meson decays from WA102 and LEAR give independent support to such relative phases\cite{CK01}
\begin{table}[h]
\begin{center}
\begin{math}
\nonumber
\bordermatrix{  & & & & \cr 
& Meson & G & s\bar{s} & n\bar{n} \cr
 & 1710: & 0.4 & 0.9 
    &  0.1 \cr 
   & 1500: & -0.6 & 0.3 & -0.7 \cr
   & 1370: & -0.7 & 0.15 &
    0.7 \cr 
  }
\end{math}
\end{center}
 \label{decaytable}
\end{table}

The fact that mass mixing and also meson decays are consistent with this set of relative phases
is interesting. The numerical values should not be taken seriously; the errors on them are probably considerable, but
the relative phases and separation of ``large, medium, small" is probably reliable. 
 
These independent analyses give a consistent interpretation of the glueball-$q\bar{q}$
mixing in the scalar channel. The challenge is how to test this?
BES and CLEO-c will soon provide over a billion $\psi$ decays giving over a thousand events
per channel in the
radiative decays $0^{++} \to \gamma (\rho;\omega;\phi)$. The ideal flavour mixing of the vector mesons
will thus ``weigh" the flavour contents of any C=+ meson
produced in $\psi \to \gamma R$. 

Some preliminary hints that there is such mixing come from the anomalous pattern of meson states $M_2$in
$\psi \to \omega/\phi + M_2$\cite{jinshan}. For an ideal flavour combination, such as $\phi = s\bar{s}$, then the
folklore is that $M_2$ will be produced via its $s\bar{s}$ content as this leads to a flavour
connected diagram. Similarly the $ \omega$ selects out $n\bar{n}$ for the $M_2$.
The test of this hypothesis has been when $M_2 \equiv 2^{++}$; this nonet consists of ideal states
$a_2(1320); f_2(1270) \equiv n\bar{n}; f_2(1525) \equiv s\bar{s}$ and therefore is rather clean
and confirms the dominance of the ``hairpin" diagram.
However, the case $M_2 = 0^{++}$ has no simple
solution. Indeed, some channels which ought to have been dominant appear even to be absent.
For example: $f_0(1370)$ has strong affinity for $\pi\pi$ and hence $n\bar{n}$ in its wavefunction
yet is not seen in $\psi \to \omega \pi\pi$. This being anomalous does not require one to suppose
that $f_0(1370)$ is $n\bar{n}$ alone; multiquark components containing non-strange flavours ought to be enough
to highlight the paradox.
One explanation could be that some other contribution leads to destructive interference. The $G$ component
in the $f_0(1370)$ wavefunction is a natural candidate for this and it has even been predicted\cite{qzfecglue} that
the strength of $b(\psi \to \phi G) \sim 10^{-3}$; if this also applies to $b(\psi \to \omega G) \sim 10^{-3}$
then the destructive interference becomes plausible. The test will be to see if the relative phases of $G$
and flavored components are in line with the observed pattern of suppressed and observed decays
$\psi \to \omega/\phi f_0$.

Below 1GeV the dynamics are controlled by the strong attractive QCD forces between
colour-spin symmetric $qq$ (or $\bar{q}\bar{q}$) pairs, for example $S=0$ $\bar{\bf 3}_F$. In flavour this equates
to attraction in $\bar{\bf 3}_F$. This leads to a nonet of low lying scalars\cite{Jaf77}. Recent
data from KLOE on $\phi \to \gamma f_0/a_0(980)$ support this picture\cite{CT02,achasov} though
the role of $K\bar{K}$ threshold in disturbing the short disance diquark clustering dynamics from the
looser molecular\cite{WI82} remains to be determined\cite{CT02,Clo03}. 
The dominant production is via the $\phi \to K\bar{K} \to K\bar{K}\gamma \to 0^{++}\gamma $ loop: 
this produces the $f_0/a_0$ via their
long range wavefunction and does not teach much about the short range QCD structure. 

These tentative ideas on diquark or molecular clustering may now be receiving support from the heavy
flavour sectors.

{\bf The X(3872): anomalous charmonium}

$B$ decays have turned out to be a novel and rich source of charmonium. Among these is
a narrow state $X(3872) \to \psi \pi\pi$. Immediately above $D\bar{D}$ threshold states can remain narrow if
 they are forbidden to decay into $D\bar{D}$. Examples are
$2^{-\pm}, 3^{--}$ and radially excited $1^{++}$ within the $c\bar{c}$; also, hybrid charmonium or
$DD^*$ molecular state.

However, each of these has problems\cite{problems}. Compared to predictions in charmonium
potential models:
$2^{--}$ and $3^{--}$ have the wrong mass and the experimental $\Gamma(\gamma 1^+)$ is too small; 
for $2^{-+}$
the $b(\psi\pi\pi)$ is expected to be small, in contrast to its visibility there; the radial $1^{++}$
is expected to have a larger $\Gamma(\gamma \psi)$ than seen; the $1^{+-}$ has a different cos$\theta$
distribution. Either standard $c\bar{c}$ theory is wrong or the
$X(3872)$ is not a simple charmonium state.

The latter is suspected to be the case, in part driven by the remarkable
coincidence between its mass and that of the threshold for $D^0D^{0*}$ which agree to better than one part in 10,000.
Refs\cite{CP04} suggest that it is a molecular or tetraquark bound state of these mesons
in S-wave; thus $1^{++}$. A particular model realisation is due to Swanson\cite{swanson}.

Observation\cite{olsen}
of the $\psi \omega$ decay supports $C=+$ and the hint that
the decay to $\psi \pi\pi$ has the $\pi\pi$
$\equiv \rho$ and not $\sigma$ support the isospin violation that the $D^0D^{0*}$
constitution would imply. Further
tests include verifying that there is no $\psi \pi^0\pi^0$: forbidden for the $\rho$
but allowed for $\sigma$. Also the hadronic decays into e.g. $K\bar{K}\pi$ will be dominated by
neutral $K^0\bar{K}^0\pi$ relative to $K^+K^-\pi$.  .

 The
 $DD^*, \psi \omega; \psi \rho$ are all effectively mass degenerate. So a mixing via
 quark exchange $D^0D^{0*} \to \psi u\bar{u}$ is driven by the energy coincidence, which
 is probably more generally true than the details of any particular model. The $u\bar{u}$ maps equally
 onto $\rho, \omega$ and so one expects $\psi\omega \sim \psi\rho$, any deviations from equality being
 a pointer to dynamical effects.
 Decays are driven by the meson components of the wavefunction while the production will be by the
 easiest route; thus seeding by the short range $c\bar{c}$ component will cause the X to be produced like
 conventional charmonium states.
 
 There may be analogues of this dynamics in the $\psi \to (K\Lambda)\bar{p}$ where the $K\Lambda$ appear
 to have $S_{11}$ baryon quantum numbers. This is another example of the S-wave hadron channels overriding the
 P-wave quark structure (in this case $qqq$): quark exchange links the $N\eta \to K\Lambda$. The $\psi
 \to \gamma p\bar{p}$ also may be showing S-wave enhancements; whether these are evidence of a bound state or
 above threshold S-wave attractions remains to be determined, though comparison with the LEAR data
 on $p\bar{p}$ annihilation just above threshold supports the bound state interpretation
 \cite{jinshan}.

{\bf Strange strange-charmed states: 2317,2460, 2635 MeV}

The $0^+,1^+$
 at 2317 and 2460MeV are lighter than the quark model had predicted, even though it had been 
 successful hitherto in this sector. One interpretation is that this is evidence for a chiral
 symmetry where the mass gap of $0^-:1^-$ equates with $0^+:1^+$\cite{chiral}. Why does this show
 up in $c\bar{s}$ where no $u,d$ chiral-friendly flavours are involved? And the axial ought to
 be the $j=1/2$ member whereas the physical states are mixtures of $j=1/2,3/2$. Thus unless one can
 argue that the 2460 is the $j=1/2$ member, the identity in the mass gaps appears a tantalising coincidence.
 The chiral relation may be applicable in the $M_Q \to \infty$ limit when $u,d$ accompany the heavy quark;
 its application in the finite mass case with $s$ is less clear.
 
 The coincidence of the masses lying just below the $DK$ and $D^*K$ thresholds has led to suggestions
 that their masses are lowered from the naive $c\bar{s}$ by a mechanism similar to that responsible
 for lowering the $f_0(980)$ and $a_0(980)$ to the vicinity of the $K\bar{K}$ threshold. The challenge now is to 
distinguish
 beetween $c\bar{s}$ and molecule interpretations. One suggestion is the radiative transition
 $1^+ \to 0^+ \gamma$. In the molecule interpretation this is driven by $D^* \to D \gamma$ which
 is known. In the $c\bar{s}$ this  branching ratio is predicted to be $1.2 \times 10^{-3}$\cite{GODFREY}.
 
 $D_s(2632)$ has anomalous branching ratios favouring $D_s\eta$ over $DK$. Attempts to 
 accomodate these within the $c\bar{s}$ picture (as a radial excitation of the $1^-$) exploiting
 nodes in the radial wavefunction are unable to do so without choosing unrealistic values of established parameters.
 Suggestions that it is a tetraquark ($cu\bar{s}\bar{u}$)  which feeds $D_s\eta$ and not $DK$
 run into problems as they also imply $D_s\pi^0$ decays. These would feed $D_s \gamma \gamma$
 but there is no sign of an enhancement at 2632 in these data. Ref.\cite{barnes} concludes
 that either our understanding of hadron decays is wrong or this state is an artefact. Its non
 observation in other experiments adds weight to this interpretation.
 
{\bf Some reflections on pentaquarks}

 If narrow width pentaquarks exist with positive parity, powerful correlations
 must arise in Strong QCD. 
 In QCD attractions are predicted between distinct flavoured pairs in net spin zero, which is the starting 
 point of two particular models\cite{JW,KL}. It has not been demonstrated how scalar diquarks form with ultra-light
 masses as required to accomodate a 1540 MeV state; their stability is an open question; their effective
 boson nature and consistency with hadron spectroscopy also are not well understood.
 But first we
 need to establish whether this state is real. I shall now review various
 features.
 
{\bf{\it Mass}}
 
 The original prediction\cite{dpp} assumed that the 1710 $N^*$ is in the $\bar{10}$ and used this to set the scale
 of mass. However $\gamma p \to p^*(\bar{10})$ is forbidden by U-spin which argues against this\cite{pdg}. The mass gap
 of 180MeV per unit of strangeness is also suspect in a quark model interpretation as it leads to a 540MeV
 spread across the $\Theta - \Xi$ multiplet even though there is only one extra strange mass in going
 from $(udud\bar{s})$ to $(usus\bar{d})$ and so a much smaller gap would be anticipated\cite{JW}.
 Beware also naive application of Gell Mann Okubo mass formulae which do not distinguish between
 $|S|$ and $S$ as one goes from $\Theta (S=+1)$ to $\Xi (S=-2)$.
 
 If the $\Theta$ should prove to be real, then no simple mapping from chiral soliton
 onto a pentaquark description seems feasible. The relation between these is more profound.
 Nonetheless a narrow state of mass $\sim$1540MeV has been claimed. But when one compares the masses
 reported in $K^+n$ versus $K^0p$ there appears to be a tantalising trend towards a difference\cite{qzfecmass}.
 Is this a hint of an explanation (see later) or that we are being fooled by poor statistics?
 
 No models successfully predict the mass; in all cases it is fitted relative to some other assumed
 measure. The original chiral soliton normalised to the 1710, as we already discussed.
 Ref.\cite{JW} assume that the Roper 1440 is the $udud\bar{d}$ (but this state is partnered by
 $\Delta(1660)$ which along with its electromagnetic and other properties, is in accord with it 
 being a radial $qqq$ excitation of the nucleon). 
 Ref\cite{KL} noted the kinematic similarity between reduced masses in their diquark-triquark model
 and the $c\bar{s}$ system. They adopted a 200MeV orbital excitation energy from the $1^--0^+(2317)$
 mass gap to realise a 1540MeV mass for the $\Theta$. However, if one makes a spin averaged mass for the
 $L=0,1$ levels, notwithstanding the questions about the low mass of the 2317, one gets nearer to a 450
 -480MeV
 energy gap and hence a $\Theta \sim$ 1800 MeV. In summary, all models appear to normalise to 
 some feature and do not naturally explain the low mass of an orbitally excited pentaquark.
 
{\bf{\it Width}}
 
 The chiral soliton model Lagrangian contains three terms with arbitrary strengths, $A,B,C$.
 Linear combinations of these can be related to the observable transition $\Delta N \pi$ and the $F/D$
 ratio for the $NN\pi$ vertex. The $\Theta N K$ vertex is then given by $g(\bar{10}) = 1-B-C$.
 We thus have one unknown $g(\Theta N K)$ described by another unknown, $C$. Ref\cite{ellis} shows the
 coupling is relatively insensitive to $F/D$ and that it is $C$ that controls $g(\Theta N K)$. In the
 non relativistic quark model it is argued\cite{dpp,ellis} that $F/D=2/3$ and the absence of $s\bar{s}$
 in the nucleon  lead to $B=1/5;C=4/5$. This has the remarkable implication that $g(\Theta N K)=0$.
 If the $\Theta$ phenomenon survives then a deeper understanding of this result and its implications
 would be welcome. It would also raise the challenge of how the $\Theta$ is strongly produced.
 
 Phenomenologically $\Gamma(\Lambda(1520) \to KN) \sim 7$MeV has been suggesttted as a
 measure for narrow widths. However this is D-wave and phase space limited: the P-wave $\Lambda(1660)$
 width is $\sim 100$MeV. Furthermore these decays require creatioon of a $q\bar{q}$;
 for the pentaquark one has $qqqqq\bar{q}$ and the challenge is to stop its decay. There are no
 indications in conventional spectroscopy underpinning a narrow width of $\sim 1$MeV
 for $\Theta$.
 
 Colour spin and flavour mismatches between $\Theta$ and $NK$ wavefunctions have been
 proposed to suppress the natural width by large factors\cite{maltman}. 
 However it is easy to override these: soft gluon exchange defeats the colour;
 spin flip costs little and flavour rearrangement can occur. Further there is colour singlet $q\bar{q}$
 in relative S-wave within the correlated models of JW and KL\cite{dudek} and their dissociation
 into $NK$ seems hard to prevent.
 
 Ref\cite{stech} suggested that overlaps of spatial wavefunctions between pentaquark
 and nucleon may lead to a suppression. However it has not been demonstrated that such is generated dynamically.
 Dudek has shown\cite{dudek} that such an effect can arise but this involves taking a non relativistic picture
 rather literally. It is also unclear how a colour $\bar{3}$ diquark is
 attracted into a tighter (smaller?) configuration than a colour singlet meson.
 
 We almost have a paradox here. The small width implies a feeble coupling to $KN$, yet something
 must couple to $\Theta$ strongly to give a normal hadronic production rate\cite{jinshan}.
  This is an enigma which we must confront.
  
 {\bf{\it Production}}
  
  We have heard several experimental limits on the hadroproduction of the $\Theta$. Some are
  not yet restrictive,e.g. the limit in $\psi \to \Theta \bar{\Theta}$ which is phase space limited
  or that in $\psi'$ decay where one can claim that there is a big price to pay for creating ten
  $q$ and $\bar{q}$. So it is possible to wriggle. However on balance the limits
  in high statistics hadroproduction appear impressive. The onus is on supporters to explain them away 
  or find a loophole.
  
  An example of such a loophole suggested here\cite{lipkin} asks why signals are in photoproduction
  but not in hadroproduction. The photon contains $s\bar{s}$ and so may be able to feed the $\bar{s}$
  needed to make $\Theta(udud\bar{s})$ in a way not so readily accessible in hadroproduction. 
  Further appeal is made to a CLAS observation that suggests that a narrow $N^*$ at $\sim 2.4$GeV 
  may be the source of $\Theta + K$. While such a dynamics can be tested by searching for
  other decay modes, forced by SU(3)\cite{lipkin}, there remain problems. CLAS see this
  (statistically insignificant) $N^*$ in $\pi$ exchange and so the photon does not appear to be
  essential: why is this object (and its progeny, the $\Theta$) not also made in hadroprodcution if
  it is made by $\pi N$? Second; while a 2.4 GeV $N^*$ may be produced in the 3-5 GeV CLAS experiment, it is
  kinematically inaccessible in the original SpRING8 experiment and in the earlier CLAS $\gamma d$. So
  the source of $\Theta$ in this latter pair would still remain to be explained.
  
  Ref\cite{CQ04} have noted that the relative photoproduction strengths of $\Theta$ and the related
  $\Sigma^+_5$ should be similar even though the scale of each individually is highly model dependent. 
  As either of these can decay into
  $K_s p$, the absence of any  $\Sigma^+_5$ signal (even after mixing with known $\Sigma^*$) 
  accompanying the claimed $\Theta$ in the HERMES data for example raises questions.
  
  Photoproduction has also been suggested as a source of kinematic peaks that 
  fake a $\Theta$\cite{dzierba}. $\gamma N \to a_2/\rho_3 N$ followed by the $K\bar{K}$ decays of
  these mesons in D/F waves give a forward-backward peaking in the c.m. along the direction of the 
  recoil nucleon and a spurious $KN$ peak.
  At first sight the experimental absence of such peaks in $K^-n$ supports for the reality of 
  the peak in $K^+n$, but it is not necessarily so simple. Charge exchange and D/F interference
  can introduce a charge asymmetry and it is claimed to
  be possible to choose phases such that a narrow peak can arise in $K^+n$ (after feeding through Monte Carlo)
  whereas broad structure would arise in $K^-n$. It has been suggested in the discussion
  sessions here that the different
  Q-values could cause a mass shift in the kinematic peak in $K^+n$ versus $K^0p$, in accord with
  the trend of the data\cite{CQ04mass}. Whether this kinematic effect is responsible may be settled when higher
  statistics data and significant Dalitz plots become available.
  
  {\bf Conclusion}
  
  Precision and variety in experiments are taking us beyond the 40 year old simple $q\bar{q}$ quark model of
  mesons. The role of strong glue in QCD is tantalising: $\psi \to \gamma \gamma V$ is a novel
  opportunity that can test the current interpretation of the mixing between scalar glueball and
  $q\bar{q}$ above 1GeV. Evidence for exotic hybrid mesons is emerging; $\chi_1 \to \pi X$ in S-wave
immediately accesses the exotic $X = 1^{-+}$ channel. The analogous $\chi_0 \to \pi X$ probes
$X = 0^{-+}$ where production of $\pi(1800)$ (a potential hybrid partner of the pion, and interesting due to its
mass degeneracy with the charmed $D$) may be studied.

Multiquark molecules are appearing. I suggest that $X(3872)$ is $1^{++}$; the $D_s(2317/2460)$
are $0^+,1^+$ shifted to below $DK/D^*K$ thresholds by dynamics analogous to those that pull the
$f_0(980)$ and $a_0(980)$ to below the $K\bar{K}$ threshold. Ways of testing this need to be
clarifyied. I suggest that the $D_s(2632)$ is an artefact: data can easily prove me wrong.

The $\Theta$, and the question of narrow width pentaquark(s), is rightly at the centre of attention.
Either the $\Theta$ is some artefact (if so, what?) or, if real, the behaviour of Strong QCD
is profound and our current model attempts will turn out to be mere tinkering.

For future historians the vote taken at this conference from around 1000 physicists
was $\sim 60\%$ believe the evidence remains inconclusive; $\sim 40\%$ believe that the $\Theta$ is
not a resonance and in the dark of the hall only a handful were convinced that
a genuine narrow resonance has been found. A vote taken a year ago at Hadron03 scored $\sim 50\%$, $25\%$
and $25\%$ respectively. Time will tell.

\end{document}